\documentclass{svmult}

\usepackage{makeidx}     
\usepackage{graphicx}    
\usepackage{multicol}    
\usepackage{amsfonts,amsmath,amssymb,latexsym}

\makeindex             

\begin{document}

\title*{Graph-based Local Elimination Algorithms in Discrete
Optimization\thanks{Research supported by FWF (Austrian Science
Funds) under the project P17948-N13.}
}
\titlerunning{Local Elimination Algorithms}
\author{Oleg Shcherbina}
\institute{Faculty of Mathematics,\\ University of Vienna\\
 Nordbergstrasse 15,  A-1090 Vienna,\\ Austria \\
\texttt{oleg.shcherbina@univie.ac.at}}
%
%
\maketitle


\section{Introduction}
\label{intro}

The use of discrete optimization
(DO)   models and algorithms makes it possible to solve many practical
problems in scheduling theory, network optimization, routing
in communication networks, facility
location, optimization in enterprise resource planning, and logistics
(in particular, in supply chain management \cite{Dolg05}).
The field of artificial intelligence
includes aspects like theorem proving,
SAT in propositional logic (see \cite {Cook71}, \cite{GPFW97}), robotics
problems, inference calculation in Bayesian networks \cite{LaSpi88},
scheduling, and others.

Many real-life DO problems contain a huge number of variables
and/or constraints that make the models intractable for currently
available DO solvers.
$NP$-hardness refers to the worst-case
complexity of problems. Recognizing problem instances that are better
(and easier for solving) than these ''worst cases'' is a rewarding
task given that better algorithms can be used for these easy cases.

Complexity theory has proved that universality and effectiveness are
contradictory requirements to algorithm complexity. But the
complexity of some class of problems decreases if the class may be
divided into subsets and the special structure of these subsets can
be used in the algorithm design.

To meet the challenge of solving large scale DO
problems (DOPs) in reasonable time, there is an
urgent need to develop new decomposition approaches \cite{BHT85},
\cite{RalG05}, \cite{Nowak}.
Large-scale DOPs are characterized not only by huge size but
also by special or sparse structure. The block form of many
DO problems is usually caused by the weak connectedness of
subsystems of real systems. One of the first examples of large
sparse linear programming (LP) problems which {\sc Dantzig} started
to study was a class of staircase LP problems for dynamic planning
 \cite{Dantzig49}, \cite{Dantzig73}, \cite{Dantzig81}.
Further examples of staircase linear programs
(see {\sc Fourer} \cite{Fourer84}) for multiperiod planning, scheduling,
and assignment, and for multistage
structural design, are included in a set of staircase test problems collected
by {\sc Ho \& Loute} \cite{HoLo81}. Staircase linear programs have also been derived
in connection with linearly constrained optimal control and stochastic
programming \cite{Wets61}. Problems of optimal hotel apartments assignment,
linear dynamic programming, labor resources allocation, control on hierarchic
structures (usually having tree-like structure), multistage integer
stochastic programming, network problems may be considered as
examples of DO problems which have staircase structure (see \cite{Shch80}, \cite{Shch83}).
The well known  SAT problem stems from classical investigations by
logicians of propositional satisfiability and has over 90 years of
history. It is possible to represent a SAT problem as a sparse DO
problem \cite{Hook}. Some applied facility location problems can be
formulated as set covering problems, set packing problems, node
packing problems \cite{Nem88}. Another class of sparse DO problems
is a production lot-sizing problem \cite{Nem88}. The frequency
assignment problem (FAP) \cite{Koster99} in mobile telephone systems communication
is a hard problem as it is closely related to the graph coloring
problem. One of the well known decomposition approaches to solving
DOPs is Lagrangean decomposition that consists of isolating sets
of constraints to obtain separate and easy to solve DO problems.
Lagrangean decomposition removes the complicating constraints from
the constraint set and inserts them into the objective function.
Most Lagrangean decomposition methods deal with special row structures.
Block angular structures with complicating variables and with complicating
variables and constraints can be decomposed using Benders
decomposition \cite{Bend62} and cross decomposition \cite{VanRoy}.
The Dantzig-Wolfe decomposition principle of LP has its equivalent
in integer programming \cite{VandS00}. This approach uses the reformulation
that gives rise to an integer master problem, whose typically large
number of variables is dealt with implicitly by using an integer
programming column generation procedure, also known as branch-and-price
algorithm \cite{BJNSV98} that allows solving large-scale DOPs in recent years.
{\sc Nemhauser} (\cite{Nem94}, p. 9) mentioned,  however, that
\begin{quote}
... the overall idea of using branch and bound with linear programming relaxation has not changed.
\end{quote}
Usually, DOPs from applications have a special structure, and the matrices
of constraints for large-scale problems have a lot of zero elements
(sparse matrices). Among decomposition approaches appropriate for
solving such problems we mention poorly known local decomposition
algorithms using the special block matrix structure of constraints
and half-forgotten nonserial dynamic programming algorithms (NSDP)
({\sc Bertele \& Brioschi} \cite{BerBri69a}, \cite{BerBri69b}, \cite{BerBri},
{\sc Dechter} \cite{Decht92}, \cite{Dechter}, \cite{Dechter2001},
\cite{Dechter03}, {\sc Hooker} \cite{Hook}) which  can
exploit sparsity in the dependency graph of a DOP and allow to compute
a solution in stages such that each of them uses results from previous
stages.

Recently, there has been growing interest in graph-based
approaches to decomposition \cite{Bodl03}; one of them is
 tree decomposition (TD).
 {\sc Courcelle} \cite{Cour90} and
{\sc Arnborg} et al. \cite{Arn91} showed that several $NP$-hard problems
posed in monadic second-order logic can be solved in polynomial time
using dynamic programming techniques on input graphs with bounded
treewidth.
Thus graph-based decomposition
approaches have gained importance.  Graph-based structural
decomposition techniques, e.g.,  nonserial dynamic programming (NSDP)
({\sc Bertele, Brioschi} \cite{BerBri},  {\sc Esogbue \& Marks} \cite{EsogMar74}, 
{\sc Hooker} \cite{Hook}, {Martelli \& Montanari} \cite{MartMont72}, 
{Mitten \& Nemhauser} \cite{Mitt63}, {\sc Neumaier \& Shcherbina} \cite{Neus}, 
{\sc Rosenthal \cite{Rosen}}, {\sc Shcherbina}
\cite{Soa07k}), {\sc Wilde \& Beightler} \cite{Wilde67}   and its modifications
(bucket elimination \cite{Dechter}, Seidel's invasion method
\cite{Sei81}), tree decomposition combined with dynamic programming
\cite{DechtPea}, \cite{BodKos07} and its variants \cite{PanGo96}, hypertree \cite{Gottlob00}
and hinge decomposition \cite{JGC94}, \cite{GJC94} are promising decomposition
approaches that allow  exploiting the
structure of  discrete problems in constraint satisfaction (CS) \cite{Fr92}
 and DO.

It is important that aforementioned methods  use just the local information
 (i.e., information about
elements of given element's neighborhood) in a process of solving
discrete problems. It is possible to propose a class of local
elimination algorithms as a general framework  that allows to
calculate some \textbf{global information} about a solution of the entire
problem using \textbf{local computations} \cite{JLO90}, \cite{LaSpi88}, \cite{SheSha86}.
Note that a main feature in aforementioned problems is
the locality of information, a definition of elements' neighborhoods
and studying them.

The use of local information (see \cite{Zhur98}, \cite{ZhurL95}, \cite{Fink65},
\cite{Soa08}, \cite{Urrut2007}) is very
important in studying complex discrete systems and in
the development of decomposition methods for solving
large sparse discrete problems; these problems
simultaneously belong to the fields of discrete optimization
\cite{Nem88}, \cite{Flo95}, \cite{PardDu}, \cite{PardWolk},  \cite{Serg03},
artificial intelligence \cite{Dechter}, \cite{Gottlob08}, \cite{Nea}, \cite{Pea4},
and databases \cite{Beeri83}. In linear algebra,
multifrontal techniques for solving sparse systems of linear
equations were developed (see \cite{Rose72}); these methods are also
of the decomposition nature. In \cite{Zhur98}, local algorithms
for computing information are
introduced. A local algorithm $A$ examines the elements in
the order specified by an ordering algorithm $A_\pi$, calculates the function
$\phi$ whose value at each step determines the form of the information
marks, and labels the element using local information about the elements
in its neighborhood. The function $\phi$ that induces the algorithm
depends on two variables: the first ranges over the set of all elements
and the second  ranges over the set of neighborhoods.
Local decomposition algorithms (see \cite{Shch80}, \cite{Shch83}) in DO problems have
a specific feature. Namely, rather than calculating predicates, they use
Bellman's optimality principle \cite{Bell} to find optimal solutions of the subproblems
corresponding to blocks of the DO problem. A step of the local
algorithm $\mathfrak{A}$ changes the neighborhood
and replaces the index $p$ by $p + 1$ (however, one can increment
the index by an arbitrary number replacing
$S_p$ by $S_{p+\rho}$; at each step of the algorithm, for every
fixed set of variables of the boundary ring, the values of
the variables of the corresponding neighborhood are stored, which
is an important difference of the local
algorithm $\mathfrak{A}$ from $A$: information about variables
in the solutions of the subproblems is stored rather than information
about the predicates. {\sc Zhuravlev} proposed to call it
\textbf{indicator information}.

Tree and branch decomposition algorithms have been
shown to be effective for DO problems like  the traveling salesman problem \cite{CS03}, frequency
assignment \cite{Koster99} etc. (see a survey paper \cite{Hicks}). 
A paper  \cite{Arn85} surveys  algorithms
that use tree decompositions. Most
of works based on tree decomposition approach only present
theoretical results \cite{Jegou_comp}, see the recent surveys
\cite{Hicks}, \cite{Soa07t}. 
Thus these methods are not yet recognized tools of operations research practitioners.

Some implementations of NSDP  are known \cite{BerBri}, \cite{Fern88},
however, generally, it remains some ''obscure'' tool for operations
research modellers. Usually, tree decomposition approaches and NSDP
are considered in the literature separately, without reference to
the close relation between these methods. We try to indicate a close
relation between these methods.

A need to solve large-scale discrete problems with special
structure using graph-based structural decomposition methods
provides the main motivation for this chapter. Here we try to
answer a number of questions about tree decomposition and NSDP
in solving DO problems. What are they? How and where can they
be applied? What consists a connection between different structural
decomposition methods, such as tree decomposition and nonserial
dynamic programming?

The aim of this paper is to provide a review of structural
decomposition methods and to give a unified framework in the
form of \textbf{local elimination algorithms} \cite{Soa08}.
We propose here the general approach which consists of viewing
a decomposition of some DO problem as being represented by a
DAG whose nodes represent subproblems that only contain local
information. The nodes are connected by arcs that represent
the dependency of the local information in the subproblems.
A subproblem that is higher in the hierarchy may use the information
(or knowledge) obtained in the dependent subproblems.

This paper is organized as follows: In section \ref{struc}
we introduce local elimination algorithms
for solving  discrete problems.
In Section \ref{dop_sec:2} we survey necessary
terminology and notions for discrete optimization problems and their graph
representations. In Section \ref{sec_nsdp:4} we consider local  variable
elimination schemes  for solving DO problems with constraints
and discuss a classification of dynamic programming (DP) computational
 procedure. Elimination Game is introduced. Application of the bucket
  elimination algorithm from CS to solving DO problems is done.
Then, in Section \ref{sec_nsdp:8}, we consider a local block elimination
scheme and related notions. As a promising abstraction
approach of solving DOPs we define clustering that merges
several variables into a single meta-variable. This allows us to create a
quotient (condensed) graph and
apply a local block elimination algorithm. In Section \ref{tree_sec:5} a tree
decomposition scheme is introduced.
Connection of  of the local elimination algorithmic schemes with tree
decomposition and a way of transforming
 the  DAG of computational local elimination procedure to tree decomposition are discussed.

\section{Local elimination algorithms for solving  discrete problems}
\label{struc}

The structure of discrete optimization problems is determined either
by the original
elements (e.g., variables) with a system of \textbf{neighborhoods}
specified for them and with the order of searching through those
elements using a \textbf{local elimination
algorithm} or by various derived structures (e.g., block or
tree-block structures). Both original and derived structures can be
 specified by the so called \textbf{structural
graph}. The \textbf{structural graph} can be the interaction graph of  the original
elements (for example, between
the variables of the problem) or the \textbf{quotient \cite{george}
(condensed \cite{Harary}) graph}. The quotient graph can
be obtained by merging a set of original elements (for example, a
subgraph) into a condensed element. The original
subset (subgraph) that formed the condensed element is called the
\textbf{detailed graph} of this element.\\
A local elimination algorithm (LEA) \cite{Soa08} eliminates local elements
of the problem's structure defined by the structural
graph by computing and storing local information about
these elements in the form of new dependencies
added to the problem.
Thus, the local elimination procedure consists of two parts:
\begin{itemize}
  \item [A.] The \textbf{forward part} eliminates elements, computes
  and stores local solutions, and finally computes the
value of the objective function;
  \item [B.] The \textbf{backward part} finds the global solution of
  the whole problem using the tables of local solutions;
the global solution gives the optimal value of the objective
function found while performing the forward
part of the procedure.
\end{itemize}
The LEA analyzes a \textbf{neighborhood}
$Nb(x)$ of the current element $x$
in the structural graph of the problem, applies an \textbf{elimination operator}
(which depends on the particular problem) to that element,
calculates the function $h(Nb(x))$
that contains \textbf{local information} about
$x$, and finds the local solution $x^*(Nb(x))$.
Next, the element $x$ is eliminated, and a clique is created from the elements of
$Nb(x)$. The elimination of elements and the creation of
cliques changes the structural graph and the neighborhoods of elements.
The backward part of the local elimination algorithm reconstructs
the solution of the whole problem
based on the local solutions $x^*(Nb(x))$.

The algorithmic scheme of the LEA is
a \textbf{DAG} in which
the vertices correspond to the local subproblems and the
edges reflect the informational dependence of the
subproblems on each other.

\section{Discrete optimization problems and their graph
representations}\label{dop_sec:2}
\subsection{Notions and definitions}
Consider a sparse DOP in the following form
\begin{equation}\label{gen1}
 F(x_{1}, x_{2}, \ldots, x_{n})=\sum_{k \in K} f_{k}(X^{k}) \rightarrow \max
\end{equation}
subject to the constraints\\
\begin{equation}\label{gen2}
 g_{i}(X_{S_i})~ R_{i}~ 0, ~~i \in M = \lbrace 1, 2, \ldots, m \rbrace,
 \end{equation}
 \begin{equation}\label{gen3}
x_{j} \in D_{j}, ~~j \in N= \lbrace 1,\ldots, n \rbrace,
\end{equation}
where\\
$X = \left\{ x_1,\dots,x_n \right\}$ is a set of discrete
variables, $X^{k} \subseteq \lbrace x_{1}, x_{2}, \ldots, x_{n}\rbrace,
k \in K=\left\{1,2,\ldots,t \right\},$ $t$ -- number of components in the
objective function, $S_{i}\subseteq \{ 1,2, \ldots, n\},~  R_{i} \in
\lbrace \leq, = ,\geq \rbrace, i \in M$; $D_j$ is a finite set of admissible
values of variable $x_j,~~j \in N$.
 Functions $f_k(X^k),~~k \in K$ are called components of the objective function
and can be defined in tabular form. We use here the notation: if $S=\{j_1,\ldots,j_q\}$
then $X_S=\{x_{j_1},\ldots,x_{j_q}\}$.\\
In order to avoid complex notation, without loss of generality, we consider further
a DOP with linear constraints and binary variables:
\begin{eqnarray}\label{goal1}
\max_X f(X) = \max_X \sum_{k \in K} f_k (X^k),
\end{eqnarray}
subject to
\begin{equation}\label{cons1}
A_{iS_i}X_{S_i} \leq b_i,~  i \in M =
\{ 1, 2, \ldots, m \},
\end{equation}
\begin{equation}\label{cons2}
x_{j}=0,1,~ j \in N = \{ 1,\ldots, n \}.
\end{equation}
We shall consider further a linear objective function (\ref{goal2}):
\begin{eqnarray}\label{goal2}
 f(x_{1}, \ldots, x_{n})= f(X)= C_N X_N=\sum_{j=1}^n c_jx_j \rightarrow \max
\end{eqnarray}
\begin{definition}\cite{BerBri}. Variables $x \in X$ and $y \in X$
interact in  DOP with constraints (we denote $x \sim y$) if
they both appear  either in the same component of the objective
function, or in the same constraint (in
other words, if variables are both either in a set $X^k$, or in
a set $X_{S_i}$).
\end{definition}
Introduce a graph representation of the DOP. Description of the DOP
structure may be done with various detailization. The structural graph
of the DOP defines which variables are in which constraints.
Structure of a DOP  can be defined either by interaction graph of initial
elements (variables in the DOP) or by various derived
structures, e.g., block structures, block-tree structures defined by
so called \textbf{quotient} (condensed or compressed \cite{Ash95},
\cite{AshLiu98}, \cite{HendRoth98}) graph.

Concrete choice of a structural graph of the DOP defines different
local elimination schemes:  nonserial dynamic programming, block decomposition,
tree decomposition etc.

If the DOP is divided into blocks corresponding to subsets of variables
(meta-variables) or to subsets of constraints (meta-constraints), then
block structure can be described by a structural quotient (condensed) graph,
whose meta-nodes correspond to subsets of the variables of
blocks and meta-edges correspond to adjacent blocks (see below, in
section \ref{meta_sec:3}).

An \textbf{interaction graph} \cite{BerBri}
(\textbf{dependency graph} by {\sc HOOKER}
\cite{Hook}) represents a structure of the DOP in a natural way.
\begin{definition} \cite{BerBri}. \textbf{Interaction graph} of the DOP
is an undirected graph $G=(X, E)$, such that
\begin{itemize}
\item[1.] Vertices $X$ of $G$ correspond to variables of the DOP;
\item[2.] Two vertices of $G$ are adjacent iff corresponding variables interact.
\end{itemize}
\end{definition}
Further, we shall use the notion of vertices that correspond
one-to-one to variables.
\begin{definition} Set of variables interacting with a variable $x \in X$
 is denoted by $Nb(x)$ and called the \textbf{neighborhood} of the
 variable $x$.
For corresponding vertices a neighborhood of a vertex $x$
 is a set of vertices of interaction graph that are linked by
 edges with $x$. Denote the latter neighborhood as $Nb_G(x)$.
\end{definition}
Introduce the following notions:
\begin{enumerate}
\item Neighborhood of a set $S \subseteq X$, $Nb_G(S) = \bigcup_{x \in S}Nb_G(x)-S $.
\item Closed neighborhood of a set $S \subseteq X$, $Nb_G[S]= Nb_G(S) \cup S$.
\end{enumerate}

\section{Local variable elimination algorithms in discrete optimization}\label{sec_nsdp:4}
\subsection{Nonserial dynamic programming and \\
classification of DP formulations}\label{sec_sdp:2}

NSDP exploits only \textbf{local computations} to solve global discrete optimization problems
and is, therefore, a particular instance of local elimination algorithm.
It appeared in 1961  with
{\sc Aris} \cite{Aris61} (see  \cite{BeiJohn65}, \cite{BerBri69a},
\cite{BerBri69b}, \cite{Mitt63}) but is
poorly known to the optimization community. This approach is used in
Artificial Intelligence under the names ''Variable Elimination'' or
''Bucket Elimination'' \cite{Dechter}. NSDP being a natural and
general decomposition approach to sparse problems solving, considers a set
of constraints and an objective function as recursively computable
function \cite{Hook}. This allows to compute a solution in stages such that each
of them uses results from previous stages.
This requires a reduced effort to find the solution.\\
Thus, the DP algorithm can be
applied to find the optimum of the entire problem by using the
connected  optimizations of the smaller DO subproblems with
the aid of existing optimization solvers.

It is worth noting that NSDP is implicit in
Hammer and Rudeanu's ''basic method'' for pseudoboolean optimization
\cite{HamRud68}. {\sc Crama, Hansen, and Jaumard} \cite{Crama90}
discovered that the basic method can exploit the structure of a
DOP with the usage of so-called \textbf{co-occurrence graph} (interaction
graph). It was found that the
complexity of the algorithm depends on  \textbf{induced width} of this graph,
which is defined for a given ordering of the variables.
Consideration of the variables in the right order may result in a
smaller induced width and faster solution \cite{Hook02}.\\
In \cite{BerBri}  mostly DO problems without constraints were
considered. Here, we consider an application of NSDP
 variable elimination algorithm to solving DO problems with
 constraints.

One of the most useful graph-based interpretations  is a
representation of computational DP procedure as a direct acyclic
graph (DAG) \cite{soa07din}  whose vertices are associated with
subproblems and whose edges express information interdependence
between subproblems.

\begin{figure}[ht]
\centering
\includegraphics[height=2cm]{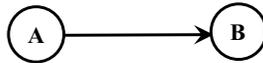}
\caption{Precedence of subproblems A and B.}
\label{AB}       
\end{figure}

\begin{figure}[h]
\centering
\includegraphics[height=6cm]{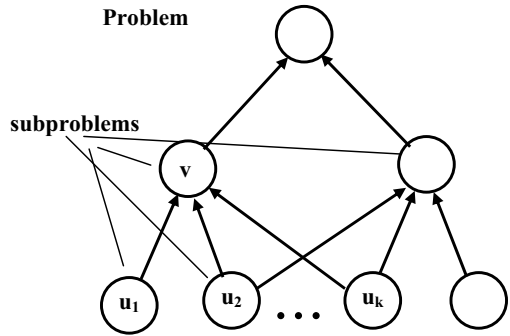}
\caption{Underlying DAG of subproblems.}
\label{dp_dag1}
\end{figure}

Every DP algorithm has an underlying
DAG structure that usually is implicit \cite{das}: the dependencies
between subproblems in a DP formulation can be represented by a DAG.
Each node in the DAG represents a subproblem. A directed edge from
node $A$ to node $B$ indicates that the solution to the subproblem
represented by node $A$ is used to compute the solution to the
subproblem represented by node $B$ (Fig.\ref{AB}). The DAG is
explicit only when we have a graph optimization problem (say, a
shortest path problem). Having nodes $u_1,\ldots,u_k$ point to $v$
means ''subproblem $v$ can only be solved once the solutions to
$u_1,\ldots, u_k$ are known'' (Fig. \ref{dp_dag1}).
Thus, the DP formulation can be described by the DAG of the computational
procedure of a DP algorithm (underlying DAG \cite{das}). {\sc Li \&
Wah}  \cite{Wah88} proposed to classify various DP
computational procedures or DP formulations on the basis of the
dependencies between subproblems from the underlying DAG.

The nodes of the DAG can be organized into
levels such that subproblems at a particular level depend only on
subproblems at previous levels. In this case, the DP procedure
(formulation) can be categorized as follows. If subproblems at all
levels depend only on the results of subproblems at the
immediately preceding levels, the procedure (formulation) is called
a \textbf{serial} DP procedure (formulation), otherwise, it is
called a \textbf{nonserial} DP procedure (formulation).
\begin{example} \label{example1} The simplest optimization problem is the  \textbf{serial}
unconstrained discrete optimization problem \cite{BerBri}
\[
\max_X f(X) = \max_X \sum_{i \in K} f_i (X^i),
\]
where $X = \left\{ x_1,\dots,x_n \right\}$ is a set of discrete
variables.
\[
K=\left\{1,2,\ldots,n-1 \right\};~~X^i=\left\{x_i,x_{i+1}\right\}.
\]
In fig. \ref{serial} it is shown an interaction graph of the serial DO
problem.
\begin{figure}[htbp]
\centering
\includegraphics [height=3cm] {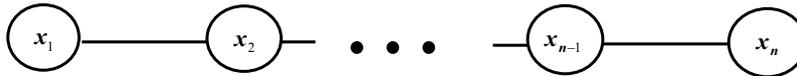}
\caption{Interaction graph for the serial formulation of unconstrained
DOP.} \label{serial}
\end{figure}
\end{example}
\subsection{Discrete optimization problem with constraints}
\label{sec_nsdp:5}

Consider the DOP (\ref{goal2}), (\ref{cons1}), (\ref{cons2}) and
suppose without loss of generality that variables are eliminated in
the order $x_1, \ldots, x_n$. Using the local variable elimination
scheme eliminate the first variable $x_1$.  $x_1$ is in a set
of constraints with the indices in $U_1$:
\[
U_{1} = \{ i \mid x_1 \in S_i \}
\]
Together with $x_1$, constraints in $U_1$ contain variables from
$Nb(x_1)$.

The following subproblem $P_1$ corresponds to the variable $x_1$  of the
DOP:
\[
h_{x_1}(Nb(x_{1}))=\max_{x_1} \{ c_1 x_1 | A_{iS_i}X_{S_i} \leq b_i,~
i \in  U_{1},~ x_{j}=0,1,~ x_{j} \in Nb[x_{1}]  \}
\]
Then the initial DOP can be transformed in the following way:
\[
\max_{x_1,\ldots, x_n} \Bigl\{ \sum C_N X_N | A_{iS_i}X_{S_i} \leq b_i,
~~i \in M, ~x_{j}=0,1,~ j \in N \Bigr\} =
\]
\[
\max_{x_2,\ldots, x_{n}} \{  C_{N-\{1\}}X_{N-\{1\}} +
h_{x_1}(Nb(x_{1})| A_{iS_i}X_{S_i} \leq b_i,~  i \in  M-U_{1},~
x_{j}=0,1,~  j = 2,\ldots, n\}
\]
The last problem has $n-1$ variables; from the initial DOP
 were excluded constraints with the indices in $U_1$ and from the
objective function the term $c_1x_1$;  there appeared a new objective
function term $h_{x_1}(Nb(x_{1}))$. Due to this fact the interaction
graph associated with the new problem is changed: a vertex $x_1$ is
eliminated and its neighbors have become connected (due to the
appearance a new term $h_{x_1}(Nb(x_{1}))$ in the objective). It can
be noted that a graph induced by vertices of $Nb(x_{1})$ is
complete, i.e. is a clique. Denote the new interaction graph $G^{1}$
and find all neighborhoods of variables in $G^{1}$. NSDP eliminates
the remaining variables one by one in an analogous manner. We have
to store tables with optimal solutions at each stage of this
process.\\ At the stage $n$ of the described process we eliminate a
variable $x_n$ and find an optimal value of the objective function.
Then a backward step of the local elimination procedure is performed using the
tables with solutions.
\subsection{Elimination game, combinatorial elimination process, and underlying DAG of the LAE
computational procedure}
\label{sec_nsdp:6}
Consider a sparse discrete optimization problem (\ref{gen1}) --- (\ref{gen3}) whose structure
is described by an undirected interaction graph  $G=(X,E)$. Solve this problem with a local
elimination algorithm (LEA). LEA uses an ordering $\alpha$ of $X$ \cite{RoTaLu76}: Given a graph
 $G=(X,E)$ an \textbf{ordering} $\alpha$ of $X$ is a bijection
 $\alpha: X \leftrightarrow \{1,2,\ldots,n\}$ where $n=|X|$.\\
$G_\alpha$ and $X_\alpha$ are correspondingly an ordered  graph and an ordered vertex set.
Sometimes the ordering will be denoted as $x_1,\ldots,x_n$, i.e. $\alpha(x_i)=i$ and
$i$ will be considered as an index of the vertex  $x_i$.

In $G_\alpha$, a \textbf{monotone neighborhood} $\overline{Nb}^{\alpha}_G(x_i)$
(\cite{Blair93}, \cite{RoTaLu76}) of $x_i \in X$ is
a set of vertices \textbf{monotonely adjacent} to a vertex $x_i$, i.e.
\[
\overline{Nb}^{\alpha}_G(x_i)=\{x_j \in Nb_G(x_i)| j > i\}.
\]

The graph $G_x$ \cite{Rose72} obtained from $G=(X,E)$ by
\begin{itemize}
\item [(i)] adding edges so that all vertices in $Nb_G(x)$ are pairwise adjacent, and
\item [(ii)] deleting $x$ and its incident edges
\end{itemize}
is the $x$--\textbf{elimination} graph of $G$.
This process is called the \textbf{elimination} of the vertex $x$.

Given an ordering  $x_1,x_2, \ldots, x_n$, the LEA proceeds in the following way:
it subsequently eliminates $x_1,x_2, \ldots, x_n$ in the current graph and computes an associated
local information about vertices from  $h_{x_i}(Nb(x_i))$ \cite{Soa08}. This
can be described by the \textbf{combinatorial elimination process} \cite{Rose72}:
\[
G_0=G,G_1, \ldots, G_{j-1}, G_{j}, \ldots, G_{n}
\]
where $G_{j}$ is the $x_j$--elimination graph of $G_{j-1}$ and $G_n = \emptyset$.

The process of interaction graph transformation corresponding to the
LEA scheme is known as \textbf{Elimination Game} which was first
introduced by {\sc{Parter}} \cite{Part} as a graph analogy of
Gaussian elimination.  The input of the elimination game is a graph
$G$ and an ordering $\alpha$ of $G$ (i.e. $\alpha(x)=i$ if $x$ is
$i$-th vertex in the ordering $\alpha$). Elimination Game according to \cite{HEKP01}
consists in the following.
 At each step $i$, the neighborhood of vertex $x_i$ is turned into a
 clique, and $x_i$ is deleted from the graph. This is referred to as
 eliminating vertex $x_i$. We obtain a graph $G_{x_i}^{(i)}$.
The filled graph $G^{+}_{\alpha}=(X,~E^{+}_{\alpha})$ is
obtained by adding to $G$ all the edges added by the algorithm.
The resulting filled graph $G_{\alpha}^{+}$ is a triangulation of
$G$ ({\sc{FULKERSON \& GROSS}} \cite{FulkGross}), i.e., a chordal
graph.

Let us introduce the notion for the \textbf{elimination tree} (etree) \cite{Liu90}.
Given a graph $G=(X,E)$ and an ordering $\alpha$, the \textbf{elimination tree}
is a directed tree $\overrightarrow{T}_{\alpha}$ that has the same vertices $X$
as $G$ and its edges are determined by a parent relation defined as follows:
the parent $x$ is the \textbf{first vertex} (according to the ordering $\alpha$) of the
monotone neighborhood $\overline{Nb}^{\alpha}_{G^{+}_{\alpha}}(x)$ of $x$ in the filled
graph $G^{+}_{\alpha}$.\\
Using the parent relation introduced above we can define a directed filled graph
$\overrightarrow{G}^{+}_{\alpha}$.\\
The underlying DAG of a local variable elimination scheme can be constructed using
Elimination  Game. At step $i$, we represent the computation
 of the function $h_{x_i}(Nb_{G_{x_{i-1}}}(x_i))$ as a node of the DAG
 (corresponding to the vertex $x_i$).
 Then, this node containing variables $(x_i, Nb_{G_{x_{i-1}}}^{(i-1)}(x_i))$ is
 linked with a first $x_j$ (accordingly to the ordering $\alpha$) which
 is in $Nb_{G_{x_{i-1}}^{(i-1)}}(x_i)$.\\
It is easy to see that the elimination tree is the DAG of the computational procedure
of the LEA.
\begin{example} \label{primer} Consider a
DOP (P) with binary variables:
\[ \begin{aligned}
&2x_1+3x_2+~x_3+5x_4+4x_5+6x_6+~x_7&\rightarrow \max\\
&3x_1+4x_2+~x_3~~~~~~~~~~~~~~~~~~~~~~~~~~&\leq 6,~~ (C_1)\\
&~~~~~~~~2x_2+3x_3+3x_4~~~~~~~~~~~~~~~~~~&\leq 5,~~ (C_2)\\
&~~~~~~~~2x_2~~~~~~~~~~~~~~~~+3x_5~~~~~~~&\leq 4,~~ (C_3)\\
&~~~~~~~~~~~~~~~~2x_3~~~~~~~~~~~~~~~~+3x_6+2x_7~&\leq 5,~~ (C_4)\\
&~x_j=0,1,~j=1,\ldots, 7.~~~~~~~~~~~~~~~~~~~~~~~~~~~~~~~~~
\end{aligned} \]
\end{example}

\begin{figure}[htbp]
\centering
\includegraphics[scale=0.7]{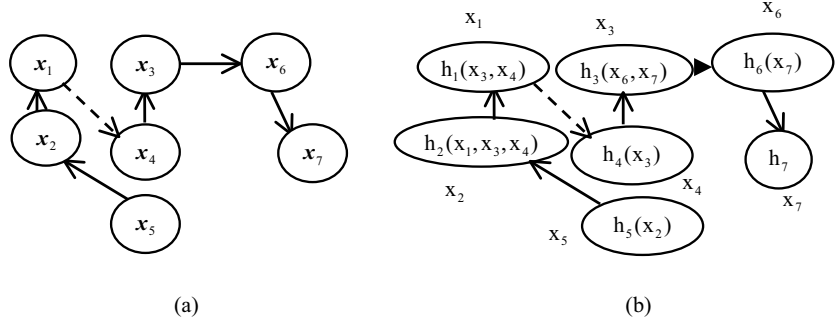} \caption{Elimination tree
of the DOP (a) Computing the information while eliminating variables
in the LEA computational procedure (b) (example \ref{primer}).} \label{InterGraph}
\end{figure}

\begin{figure}[htbp]
  \centering
 \includegraphics [height=14 cm] {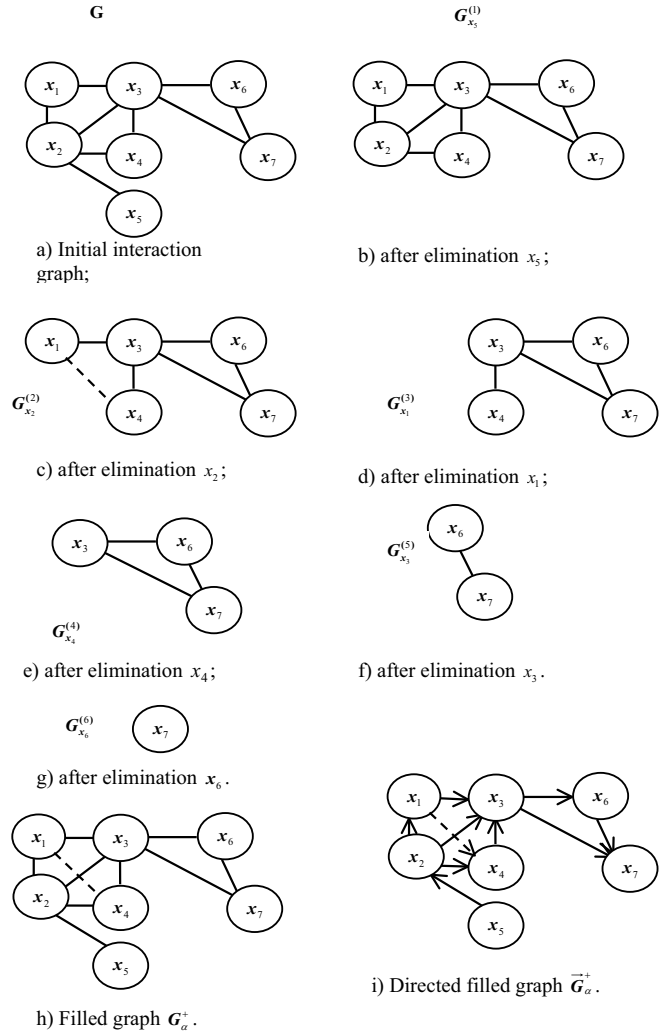} \caption{Elimination Game. Fill-in is
represented by dashed lines} \label{ElimGame}
\end{figure}

The interaction graph  is shown in
Fig. \ref{InterGraph} (a). Elimination Game results and graphs
$G_{x_{i}}^{(i)}$ are in Fig. \ref{ElimGame}. Associated
 underlying DAG of NSDP procedure for the variable ordering $\{x_5,x_2,x_1,x_4,x_3,x_6,x_7\}$
 is shown in Fig. \ref{InterGraph} (b).
\subsection{Bucket elimination}
Bucket elimination (BE)  is proposed in \cite{Dechter} as a version
of NSDP for solving CSPs. Now, we consider a modification of the BE
algorithm  for solving DOPs. The BE algorithm works
as follows: Assume we are given an order $x_1,\ldots,x_n$ of the
variables of the DOP. BE starts by creating $n$ ''buckets'', one for
each variable $x_j$.   BE algorithm uses as input ordered set of
variables and a set of constraints. To each variable $x_j$ is
corresponded a bucket $\Sigma^{(x_j)}$, i.e., a set of constraints
and components of objective function built as follows: In the bucket
$\Sigma^{(x_j)}$ of variable $x_j$ we put all constraints that
contain $x_j$ but do not contain any variable having a higher index.
We now iterate on $j$ from $n$ to 1, eliminating one bucket at a
time. Algorithm finds new components of the objective applying so
called ''elimination operator'' (in our case the latter consists on
solving associated DO subproblems) to all constraints and components
of the objective function of the bucket under consideration. New
components of the objective function reflecting an impact of
variable $x_j$ on the rest part of the DO problem, are located in
corresponding lower buckets.\\
Consider an application of BE to solving the DOP with
constraints from Example \ref{primer}. We use an elimination ordering
$\alpha:~\{x_5,x_2,(x_1,x_4),x_3,(x_6,x_7)\}$. Variables $(x_1,x_4)$
shall be eliminated in block since they are indistinguishable. Build buckets
(subsets of constraints) beginning from last (due order $\alpha$)
block $(x_6, x_7)$. A bucket $\Sigma^{(x_6, x_7)}$  includes  all constraints
of the DOP containing the variables $x_6,~ x_7$, i.e., the bucket
$\Sigma^{(x_6, x_7)}$ consists of constraint $C_4$: $\Sigma^{(x_6, x_7)}=\{C_4\}.$
Similarly: $\Sigma^{(x_3)}=\{C_1,C_2\},
~~\Sigma^{(x_1, x_4)}=\emptyset,~~ \Sigma^{(x_2)}=\{C_3\},
~~ \Sigma^{(x_5)}=\emptyset $.\\
We solve a DO subproblem associated with the bucket $\Sigma^{(x_6, x_7)}$:\\
For each binary assignment $x_3$, we compute values $x_6, x_7$ such that
\[ h_{x_6,x_7}(x_3)=\max_{x_6,x_7}\{6x_6+x_7\mid 2x_3+3x_6+2x_7 \leq 5, x_j\in
 \{0,1\}\}.\]
\begin{center}
\parbox{50mm}{
Table 1. \\{\bf Calculation of $h_{x_6,x_7}(x_3)$}\\[1ex]
\begin{tabular}{{|c|c|cc|}}
\hline\noalign{\smallskip}
$x_3$  & $h_{x_6,x_7}$  & $x_6^*$ & $x_7^*$ \\
\noalign{\smallskip}\hline\noalign{\smallskip}
0 &7&   1 &   1 \\
1 &6&   1 &   0 \\
\noalign{\smallskip}\hline
\end{tabular}}
\qquad
\parbox{50mm}{
Table 2. \\{\bf Calculation of $ h_{x_3}(x_1,x_2,x_4)$}\\[1ex]
\begin{tabular}{|ccc|c|c|}
\hline\noalign{\smallskip}
$x_1$ & $x_2$ & $x_4$ & $h_{x_3}$ & $x_3^*$ \\
\noalign{\smallskip}\hline\noalign{\smallskip}
0 & 0 &0&   7 &   1 \\
0 & 0 &1&   7 &   0 \\
0 & 1 &0&   7 &   1 \\
0 & 1 &1&   7 &   0 \\
1 & 0 &0&   7 &   1 \\
1 & 0 &1&   7 &   0 \\
1 & 1 &0&   - &   - \\
1 & 1 &1&   - &   - \\
\noalign{\smallskip}\hline
\end{tabular}}
\end{center}
The function $h_{x_6,x_7}(x_3)$ is placed in the bucket
$\Sigma^{(x_3)}$. Consider the DO subproblem associated with this
bucket
\[
\begin{aligned}
& h_{x_3}(x_1,x_2,x_4)= \max_{x_3} \left[~x_3+h_{x_6,x_7}(x_3)\right]\\
&~~~~~~~~~~~~~~~~~~~3x_1+4x_2+~x_3~~~~~~~~~~\leq 6,\\
&~~~~~~~~~~~~~~~~~~~~~~~~~~~2x_2+3x_3+3x_4~~\leq 5,\\
&~~~~~~~~~~~~~~~~~~~x_j=0,1,~j=1,2,3,4.~~~~~~~~~~~~~~~~~~~~~~~~~~~~~~~~~\\
\end{aligned}
\]
We place the function $ h_{x_3}(x_1,x_2,x_4)$  in the bucket
$\Sigma^{(x_1, x_4)}$ and solve the problem
\[
h_{x_1,x_4}(x_2)=\max_{x_1, x_4}\{2x_1+5x_4+ h_{x_3}(x_1,x_2,x_4) \mid x_j\in
\{0,1\}\}.
\]
Build the corresponding table 3.

Function $h_{x_1,x_4}(x_2)$ is placed in the bucket $\Sigma^{(x_2)}$.
A new DO subproblem left to be solved
\[
h_{x_2}(x_5)=\max_{x_2}\{3x_2+ h_{x_1,x_4}(x_2) \mid 2x_2+3x_5 \leq 4,
x_j \in \{0,1\}\}
\]
\begin{center}
\parbox{50mm}{
Table 3. \\{\bf Calculation of $h_{x_1,x_4}(x_2)$}\\[1ex]
\begin{tabular}{|c|c|cc|}
\hline\noalign{\smallskip}
$x_2$ & $h_{x_1,x_4}$& $x_1^*$& $x_4^*$ \\
\noalign{\smallskip}\hline\noalign{\smallskip}
0 &   14 &   1 & 1 \\
1 &   12 &   0 & 1 \\
\noalign{\smallskip}\hline
\end{tabular}}
\qquad
\parbox{50mm}{
Table 4.\\ {\bf Calculation of $h_{x_2}(x_5)$}\\[1ex]
\begin{tabular}{|c|c|c|}
\hline\noalign{\smallskip}
$x_5$ & $h_{x_2}$ & $x_2^*$ \\
\noalign{\smallskip}\hline\noalign{\smallskip}
0&   15 &   1 \\
 1&   14 &   0 \\
\noalign{\smallskip}\hline
\end{tabular}}
\end{center}
Place $h_{x_2}(x_5)$ in the last bucket $\Sigma^{(x_5)}$.  The new subproblem
is:
\[
h_{x_5}=\max_{x_5} \{4x_5+ h_{x_2}(x_5) \mid x_j \in \{0,1\}\},
\]
its solution is $h_5=18,~ x_5^*=1$ and the maximal objective value
is 18.

To find the optimal values
of the variables, it is necessary to do backward step of the BE
procedure: from the last table 4 using $x_5=1$ we have
$x_2^*=0$. Considering the table 3 we have for $x_2=0: x_1^*=1,~
x_4^*=1$. From the table 2: $x_1=1,~ x_2=0, ~x_4=1 \Rightarrow x_3^*=0$.
Table 1: $x_3=0 \Rightarrow x_6^*=1,~x_7^*=1$.\\
The solution is (1,~ 0,~ 0,~ 1,~ 1,~ 1,~ 1), optimal objective value
is 18.

\section{Block local  elimination scheme}
\label{sec_nsdp:8}
\subsection{Partitions, clustering, and quotient graphs}
\label{meta_sec:3}
The local elimination  procedure can be applied to elimination
of not only separate variables but also to sets of variables and can use the
so called ''elimination of variables in blocks'' (\cite{BerBri},
\cite{Shch83}), which allows to eliminate several variables in
block. Local decomposition algorithm \cite{Shch83} actually
implements the  local block elimination algorithm. If the DOP is divided into
blocks corresponding to subsets of variables
(meta-variables), then block structure can be described with the aid
of a structural condensed graph whose meta-nodes correspond to
subsets of the variables or blocks and meta-edges correspond to
adjacent blocks.

Applying the method of merging variables into meta-variables allows to
obtain \textbf{condensed} or meta-DOPs which have a simpler
structure. If the resulting meta-DOP has a nice
structure (e.g., a tree structure) then it can be solved
efficiently.

The structural graph of the meta-DOP is
obtained by collapsing merged nodes into a single meta-node and
connecting the meta-node with all nodes that were adjacent with some
of the merged nodes. Such a graph usually is called a quotient graph.\\
An \textbf{ordered} partition of a set $X$ is a decomposition of $X$
into ordered sequence of pairwise
disjoint nonempty subsets whose union is all of $X$.\\
Partitioning is a fundamental operation on graphs. One variant of it
is to partition the vertex set $X$ to three sets $X=U \cup S \cup
W$, such that $U$ and $W$ are balanced, meaning that neither of them
is too small, and $S$ is small. Removing $S$ along with all edges
incident on it separates the graph into two connected components.
$S$ is called a \textbf{separator}. In general, graph partitioning
is $NP$-hard. Since graph partitioning is difficult in general,
there is a need for approximation algorithms. A popular algorithm in
this respect is MeTiS \cite{Metis}, which has a good implementation
available in the public domain.

Taking advantage of \textbf{indistinguishable} variables (two variables are
indis-\\tinguishable if they have the same closed neighborhood
 \cite{Amestoy96},  \cite{Ash95}, \cite{HendRoth98}, \cite{AshLiu98})
 it is possible
 to compute a \textbf{quotient (condensed) graph} which is
 formed by merging all vertices with the same neighborhoods into a
 single meta-node. Let $\mathbf{x}$ be a \textbf{block} of a graph $G$ \cite{Arn87},
 i.e., a maximal set of  indistinguishable with $v$ vertices. Clearly,
 the blocks of $G$ partition $X$ since indistinguishability is an \textbf{equivalence
 relation} defined on the original vertices.

An equivalence relation on a set induces a partition on it, and also
any partition induces an \textbf{equivalence relation}. Given a graph $\Gamma
= (X, E)$, let $\mathbf{X}$ be a partition on the vertex set $X$:
\[ \mathbf{X} =
\{ \mathbf{x_1},\mathbf{x_2}, \ldots, \mathbf{x_m} \}.
\]
 That is, $\cup_{i=1}^{m} \mathbf{x_i} = X$
and $\mathbf{x_i} \cap \mathbf{x_k} = \emptyset$ for $i \ne k$. We define the
\textbf{quotient graph} of $G$ with respect to the partition
$\mathbf{X}$ to be the graph
\[
G / \mathbf{X}  = (\mathbf{X}, \mathcal{E}),
\]
where $(\mathbf{x_i},~\mathbf{x_k}) \in \mathcal{E}$ if and only if $Nb_G(\mathbf{x_i}) \cap  \mathbf{x_k}
\ne \emptyset$.

The quotient graph $\mathbf{G}(\mathbf{X},\mathcal{E})$
  is an equivalent  representation of the interaction graph $G(X,E)$,
  where $\mathbf{X}$ is a set of blocks (or indistinguishable sets of vertices),
 and $\mathcal{E} \subseteq \mathbf{X} \times \mathbf{X}$ be the edges defined
on $\mathbf{X}$. A \textbf{local block  elimination} scheme
is one in which the vertices of each block are eliminated
contiguously \cite{Arn87}. As an application of a
clustering technique we consider below a block local elimination
 procedure \cite{BerBri} where the elimination of the block
(i.e., a subset of variables) can
be seen as the merging of its variables into a meta-variable.

The merges done define a so called \textbf{synthesis tree}
\cite{WF99} on the variables.
\begin{definition}
A synthesis tree of an initial  DOP $P$ is a tree whose leaves
correspond to the variables of the initial DOP $P$, and where each
intermediate node is a meta-variable corresponding to the
combination of its children nodes.
\end{definition}
Using the synthesis tree it is possible to ''decode'' meta-variables
and find the solution of the initial DOP.

Consider an ordered partition $\mathbf{X}$ of the set $X$ of the variables into blocks:
\[
\mathbf{X}=(\mathbf{x_1},\ldots,\mathbf{x_p}),~~p \leq n,
\]
where $\mathbf{x_l}=X_{K_l}$ ($K_l$ is a set of indices corresponding to $\mathbf{x_l},~l=1,\ldots,p$).
For this ordered partition $\mathbf{X}$, the DOP P: (\ref{goal2}), (\ref{cons1}),
(\ref{cons2}) can be solved by the LEA using \textbf{quotient interaction graph} $\mathbf{G}$.

\textbf{A. Forward part}

Consider first the block $\mathbf{x_1}$. Then
\[
\max_{X} \{C_N X_N|A_{iS_i}X_{S_i} \leq b_i,~  i \in  M,~
x_{j}=0,1,~ j \in N\} =
\]
\[
\max_{X_{K_2},\ldots,X_{K_p}} \{C_{N-K_1} X_{N-K_1}+
h_1(Nb(X_{K_1})|A_{iS_i}X_{S_i} \leq b_i,~ i \in M-U_1,\]
\[ x_{j}=0,1,~j \in N-K_1\}
\]
 where $U_1=\{i:S_i \cap K_1 \ne \emptyset\}$ and
\[
h_1(Nb(X_{K_1})) = \max_{X_{K_1}} \{C_{K_1}
X_{K_1}|A_{iS_i}X_{S_i} \leq b_i,~  i \in U_1,~ x_{j}=0,1,~ x_j \in Nb[\mathbf{x_1}]\}.
\]
The first step of the local block  elimination procedure consists of
solving, using complete enumeration of $X_{K_1}$, the following
optimization problem
\begin{equation}\label{h_1}
h_1(Nb(X_{K_1})) = \max_{X_{K_1}} \{C_{K_1} X_{K_1}|A_{iS_i}X_{S_i}
\leq b_i,~  i \in U_1,~ x_{j}=0,1,~ x_j \in Nb[\mathbf{x_1}]\},
\end{equation}
and storing the optimal local solutions $X_{K_1}$ as a function of
the neighborhood $of X_{K_1}$, i.e., $X_{K_1}^*(Nb(X_{K_1}))$.

The maximization of $f(X)$ over all feasible assignments
$Nb(X_{K_1})$, is called the \textbf{elimination of the block} (or
meta-variable) $X_{K_1}$. The optimization problem left after the
elimination of $X_{K_1}$ is:
\[
\max_{X-X_{K_1}} \{C_{N-K_1} X_{N-K_1}+
h_1(Nb(X_{K_1}))|A_{iS_i}X_{S_i} \leq b_i,~ i \in M-U_1,\]
\[
x_{j}=0,1,~ j \in N-K_1 \}.
\]

Note that it has the same form as the original problem, and the
tabular function  $h_1(Nb(X_{K_1}))$ may be considered as a new
component of the modified objective function. Subsequently, the same
procedure may be applied to the elimination of the blocks --
meta-variables $\mathbf{x_2}=X_{K_2},\ldots,\mathbf{x_p}=X_{K_p}$,
in turn. At each step $j$
the new component $h_{\mathbf{x_j}}$ and optimal local solutions $X^{*}_{K_j}$
are stored as functions of $Nb(X_{K_j} \mid X_{K_1}, \ldots,
X_{K_{j-1}})$, i.e., the set of variables interacting with at least
one variable of $X_{K_j}$ in the current problem, obtained from the
original problem by the elimination of $X_{K_1}, \ldots,
X_{K_{j-1}}$. Since the set $Nb(X_{K_{p}} \mid X_{K_{1}}, \ldots,
X_{K_{p-1}})$ is empty, the elimination of $X_{K_{p}}$ yields the
optimal value of objective $f(X)$.

\textbf{B. Backward part.}

This part of the procedure consists of the consecutive choice of
$X^{*}_{K_p}$, $X^{*}_{K_{p-1}},\ldots, X^{*}_{K_{1}}$, i.e., the
optimal local solutions from the stored tables $ X^{*}_{K_{1}}
(Nb(X_{K_{1}})), X^{*}_{K_{2}}(Nb(X_{K_{2}} \mid X_{K_{1}})),
\ldots, X^{*}_{K_{p}} \mid X_{K_{p-1}},\ldots, X_{K_{1}}$.

\textbf{ Block elimination game and underlying DAG}

It is possible to extend EG to the case of the block elimination. The
input of extended EG is an initial interaction graph $G$ and a
partition $\mathbf{X}=\{\mathbf{x_1},\ldots,\mathbf{x_p}\}$ of vertices of $G$.
At each step $\nu$ ($1 \leq \nu \leq p$) of EG, the neighborhood
$Nb(\mathbf{x_\nu})$ of $\mathbf{x_\nu}$ is turned into a clique, and  $\mathbf{x_\nu}$ is
deleted from the graph $G$. The filled graph $G_{\mathbf{X}}^{+}=(X,E^{+})$
 is obtained by adding to $G$ all the edges added by the algorithm.
The resulting filled graph $G_{\mathbf{X}}^{+}$ is a triangulation
of $G$, i.e., a chordal graph \cite{Arn91}.

Underlying DAG of the local block elimination procedure contains nodes corresponding
to computing of functions $h_{\mathbf{x_i}}(Nb_{G_{\mathbf{X}}^{(i-1)}}(\mathbf{x_i}))$
and is a \textbf{generalized elimination tree}.

 \begin{example}
\label{ch2:ex_nonser_unconstr_block}

\textbf{ Local block elimination for unconstrained DOP.}

Consider an unconstrained DOP
 \[ \max_X
[f_1(x_1,x_2,x_3)+f_2(x_2,x_3,x_4)+f_3(x_2,x_5)+f_4(x_3,x_6,x_7)],
\]
where
\[ X=(x_1,x_2,x_3,x_4,x_5,x_6,x_7) \]
and
functions $f_1,~f_2,~f_3,~f_4$ are given in the following tables.
\begin{center}
\parbox{50mm}{
Table 5. {\bf $f_1$}\\[1ex]
\begin{tabular}{|ccc|c|}
\hline\noalign{\smallskip}
$x_1$ & $x_2$ & $x_3$ & $f_1$  \\
\noalign{\smallskip}\hline\noalign{\smallskip}
0 & 0 & 0 & 2 \\
0 & 0 & 1 & 3 \\
0 & 1 & 0 & 4 \\
0 & 1 & 1 & 0 \\
1 & 0 & 0 & 5 \\
1 & 0 & 1 & 2 \\
1 & 1 & 0 & 4 \\
1 & 1 & 1 & 1 \\
\noalign{\smallskip}\hline
\end{tabular}}
\qquad
\parbox{50mm}{
Table 6. {\bf $f_2$}\\[1ex]
\begin{tabular}{|ccc|c|}
\hline\noalign{\smallskip}
$x_1$ & $x_2$ & $x_3$ & $f_2$  \\
\noalign{\smallskip}\hline\noalign{\smallskip}
0 & 0 & 0 & 3 \\
0 & 0 & 1 & 1 \\
0 & 1 & 0 & 5 \\
0 & 1 & 1 & 2 \\
1 & 0 & 0 & 4 \\
1 & 0 & 1 & 1 \\
1 & 1 & 0 & 3 \\
1 & 1 & 1 & 0 \\
\noalign{\smallskip}\hline
\end{tabular}}
\end{center}

\bigskip

\begin{center}
\parbox[t]{30mm}{
Table 7. {\bf $f_3$}\\[1ex]
\begin{tabular}{|cc|c|}
\hline\noalign{\smallskip}
 $x_2$ & $x_5$ & $f_3$  \\
\noalign{\smallskip}\hline\noalign{\smallskip}
0 & 0 & 6 \\
0 & 1 & 2 \\
1 & 0 & 4 \\
1 & 1 & 5 \\
\noalign{\smallskip}\hline
\end{tabular}}
\qquad
\parbox[t]{50mm}{
Table 8. {\bf $f_4$}\\[1ex]
\begin{tabular}{|ccc|c|}
\hline\noalign{\smallskip}
$x_3$ & $x_6$ & $x_7$ & $f_4$  \\
\noalign{\smallskip}\hline\noalign{\smallskip}
0 & 0 & 0 & 5 \\
0 & 0 & 1 & 2 \\
0 & 1 & 0 & 3 \\
0 & 1 & 1 & 4 \\
1 & 0 & 0 & 2 \\
1 & 0 & 1 & 1 \\
1 & 1 & 0 & 3 \\
1 & 1 & 1 & 6 \\
\noalign{\smallskip}\hline
\end{tabular}}
\end{center}
Consider an ordered
partition of the variables of the set $Õ$ into blocks: \[
\mathbf{x_1}=\{x_5\},~\mathbf{x_2}=\{x_1,x_2,x_4\},~ \mathbf{x_3}=\{x_6,x_7\},~ \mathbf{x_4}=\{x_3\}. \]
Interaction graph for this problem is the same as in Fig. \ref{InterGraph} (a).

For the ordered partition $\mathbf{X}=\{\mathbf{x_1},~\mathbf{x_2},~ \mathbf{x_3},~ \mathbf{x_4}\}$, this
unconstrained DO problem may be solved by the LEA. Initial
interaction graph with partition presented by dashed lines is shown in
Fig. \ref{BlockGraph} (a), quotient interaction graph is in Fig. \ref{BlockGraph} (b),
and the DAG of the block local elimination computational procedure is shown in Fig. \ref{DAG_BlockGraph}.
\begin{figure}[htbp]
\centering
\includegraphics[scale=0.7]{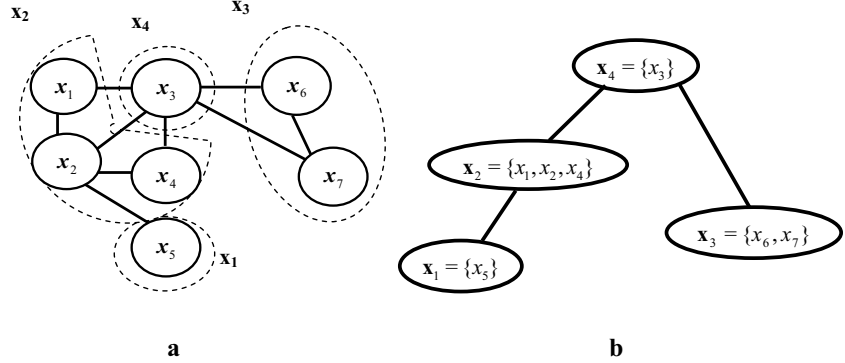}
\caption{Interaction graph of the DOP with partition (dashed) (a) and quotient interaction graph (b)
(example \ref{ch2:ex_nonser_unconstr_block}).}\label{BlockGraph}
\end{figure}
\begin{figure}[htbp]
\centering
\includegraphics[scale=0.7]{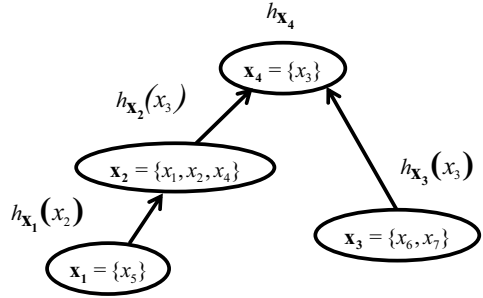}
\caption{The DAG (generalized elimination tree) of the local block  elimination computational procedure
for the DO problem (example \ref{ch2:ex_nonser_unconstr_block}).}\label{DAG_BlockGraph}
\end{figure}

\textbf{A. Forward part}

Consider first the block $\mathbf{x_1}=\{x_5\}$. Then $Nb(\mathbf{x_1})=\{x_2\}$.
Solve using complete enumeration the following optimization
problem \[ h_{\mathbf{x_1}}(Nb(\mathbf{x_1})) = \max_{x_5} f_3 (x_2,x_5), \] and store
the optimal local solutions $\mathbf{x_1}$ as a function of a neighborhood,
i.e., $\mathbf{x_1}^*(Nb(\mathbf{x_1}))$.

Eliminate the block $\mathbf{x_1}$ and consider the block $\mathbf{x_2}=\{x_1,x_2,x_4\}$.
$Nb(\mathbf{x_2})=\{x_3\}$. Now the problem to be solved is
\[
h_{\mathbf{x_2}}(x_3)=\max_{x_1,x_2,x_4}\{h_{\mathbf{x_1}}(x_2)+f_1(x_1,x_2,x_3)+f_2(x_2,x_3,
x_4)\}.
\]
Build the corresponding table 10. 
\begin{center}
\parbox[t]{50mm}{
Table 9. \\{\bf Calculation of $h_{\mathbf{x_1}}(x_2)$}\\[1ex]
\begin{tabular}{|c|c|c|}
\hline\noalign{\smallskip}
 $x_2$&   $h_{\mathbf{x_1}}(x_2)$ & $x_5^*$ \\
\noalign{\smallskip}\hline\noalign{\smallskip}
0 &   6 &   0 \\
1 &   5 &   1 \\
\noalign{\smallskip}\hline
\end{tabular}}
\qquad
\parbox[t]{50mm}{
Table 10.\\ {\bf Calculation of $h_{\mathbf{x_2}}(x_3)$}\\[1ex]
\begin{tabular}{|c|c|ccc|}
\hline\noalign{\smallskip}
 $x_3$& $h_{\mathbf{x_2}}(x_3)$ & $x_1^*$ & $x_2^*$ & $x_4^*$\\
\noalign{\smallskip}\hline\noalign{\smallskip}
0 &14&   1 &   0 & 0 \\
1 &14&   0 &   0 & 0 \\
\noalign{\smallskip}\hline
\end{tabular}}
\end{center}
Eliminate the block $\mathbf{x_2}$ and consider the block
$\mathbf{x_3}=\{x_6,x_7\}$. The neighbor of $\mathbf{x_3}$ is $x_3$:
$Nb(\mathbf{x_3})=\{x_3\}$. Solve the DOP containing $x_3$:
\[
h_{\mathbf{x_3}}(x_3)=\max_{x_6,x_7}\{f_4(x_3,x_6,x_7), x_j\in \{0,1\}\}
\]
and build the table 11.
\begin{center}
\parbox[t]{50mm}{
Table 11. \\{\bf Calculation of $h_{\mathbf{x_3}}(x_3)$}\\[1ex]
\begin{tabular}{|c|c|cc|} \hline
 $x_3$&   $h_{\mathbf{x_3}}(x_3)$ & $x_6^*$ & $x_7^*$\\
\hline 0 &   5 &   0 & 0 \\ 1 &   6 &   0 & 1 \\ \hline
\end{tabular}}
 \end{center}
Eliminate the block $\mathbf{x_3}$  and consider the block
$\mathbf{x_4}=\{x_3\}$. $Nb(\mathbf{x_4})=\emptyset$. Solve the DOP:\\
\[
h_{\mathbf{x_4}}=\max_{x_3}\{h_{\mathbf{x_2}}(x_3)+h_{\mathbf{x_3}}(x_3), x_j\in \{0,1\}\}=20,
\] where $x_3^*=1$.

 \textbf{B. Backward part.}

Consecutively find $\mathbf{x_3}^*, \mathbf{x_2}^*, \mathbf{x_1}^*$, i.e., the optimal
local solutions from the stored tables 11, 10, 9:\\
$ x_3^*=1 \Rightarrow x_6^*=1,~ x_7^*=1$ (table 11);\\
$ x_3^*=1 \Rightarrow x_1^*=0,~ x_2^*=0,~ x_4^*=0$ (table 10);
$ x_2^*=0 \Rightarrow x_5^*=0$ (table 9).\\
We found the optimal solution to be $(0,~0,~1,~0,~0,~1,~1)$, the maximum objective value
is 20.
\end{example}
\begin{example}\label{ch2:ex_nonser_constr_block}

\textbf{ Local block elimination for constrained  DOP}

Consider the DOP from example \ref{primer} and an ordered partition of the variables of
the set $Õ$ into blocks:
\[
\mathbf{x_1}=\{x_5\},~\mathbf{x_2}=\{x_1,x_2,x_4\},~
\mathbf{x_3}=\{x_6,x_7\},~ \mathbf{x_4}=\{x_3\}.
 \]
For the ordered partition $\{\mathbf{x_1},~\mathbf{x_2},~ \mathbf{x_3},~ \mathbf{x_4}\}$, this
constrained optimization problem may be solved by the LEA.

\textbf{A. Forward part}

Consider first the block $\mathbf{x_1}=\{x_5\}$. Then $Nb(\mathbf{x_1})=\{x_2\}$.
Solve the following problem containing $x_5$ in the objective and
the constraints:\\
\[
h_{\mathbf{x_1}}(Nb(\mathbf{x_1}))=\max_{x_5}\{4x_5\mid
2x_2+3x_5 \leq 4, x_j\in \{0,1\}\}
\]
and store the optimal local
solutions $\mathbf{x_1}$ as a function of a neighborhood, i.e.,
$\mathbf{x_1}^*(Nb(\mathbf{x_1}))$.
Eliminate the block $\mathbf{x_1}$.
and consider the block $\mathbf{x_2}=\{x_1,x_2,x_4\}$.
$Nb(\mathbf{x_2})=\{x_3\}$. Now the problem to be solved is
\[
\begin{aligned}
& h_{\mathbf{x_2}}(x_3)=\max_{x_1,x_2,x_4} \{h_{\mathbf{x_1}}(x_2)+
2x_1+3x_2+5x_4\} \\
\hbox{subject to}\\
&~~~3x_1+4x_2+~x_3~~~~~~~~~~\leq 6,\\
&~~~~~~~~~~~~2x_2+3x_3+3x_4~\leq 5,\\
&~~~~~~~~~~~~~~~~~~~~~~~x_j=0,1,~j=1,2,3,4.~~~~~~~~~~~~~~~~~~~~~~~~~~~~~~~~
\end{aligned}
\]
Build the corresponding table 13.
\begin{center}
\parbox[t]{40mm}{
Table 12. \\{\bf Calculation of $h_{\mathbf{x_1}}(x_2)$}\\[1ex]
\begin{tabular}{|c|c|c|}
\hline\noalign{\smallskip}
$x_2$&   $h_{\mathbf{x_1}}(x_2)$ & $x_5^*$ \\
\noalign{\smallskip}\hline\noalign{\smallskip}
0 &   4 &   1 \\
1 &   0 &   0 \\
\noalign{\smallskip}\hline
\end{tabular}}
\qquad
\parbox[t]{50mm}{
Table 13. \\{\bf Calculation of $h_{\mathbf{x_2}}(x_3)$}\\[1ex]
\begin{tabular}{|c|c|ccc|}
\hline\noalign{\smallskip}
 $x_3$& $h_{\mathbf{x_2}}(x_3)$ & $x_1^*$ & $x_2^*$ & $x_4^*$\\
\noalign{\smallskip}\hline\noalign{\smallskip}
0 &11&   1 &   0 & 1 \\
1 &6&   1&   0 & 0 \\
\noalign{\smallskip}\hline
\end{tabular}}
\end{center}
Eliminate the block $\mathbf{x_2}$ and consider the block
$\mathbf{x_3}=\{x_6,x_7\}$.
The neighbor of $\mathbf{x_3}$ is $x_3$:
$Nb(\mathbf{x_3})=\{x_3\}$. Solve the DOP containing $x_3$:\\
$h_{\mathbf{x_3}}(x_3)=\max_{x_6,x_7}\{h_{\mathbf{x_2}}+x_3+6x_6+x_7~|~2x_3+3x_6+2x_7
\leq 5, x_j\in \{0,1\}\}$\\ and build the table 14.
\begin{center}
\parbox[t]{40mm}{
Table 14. \\{\bf Calculation
of $h_{\mathbf{x_3}}(x_3)$}\\[1ex]
\begin{tabular}{|c|c|cc|} \hline
 $x_3$&   $h_{\mathbf{x_3}}(x_3)$ & $x_6^*$ & $x_7^*$\\
\hline
0 &   18 &   1 & 1 \\
1 &   12 &   1 & 0 \\ \hline
\end{tabular}}
\end{center}
Eliminate the block $\mathbf{x_3}$ and consider the block
$\mathbf{x_4}=\{x_3\}$. $Nb(\mathbf{x_4})=\emptyset$. Solve the DOP:
\[
h_{\mathbf{x_4}}=\max_{x_3}\{h_{\mathbf{x_3}}(x_3), x_j\in \{0,1\}\}=18,
\] where $x_3^*=0$.

 \textbf{B. Backward part.}

Consecutively find $\mathbf{x_3}^*, \mathbf{x_2}^*, \mathbf{x_1}^*$, i.e., the optimal
local solutions from the stored tables 14, 13, 12.
$ x_3^*=0
\Rightarrow x_6^*=1,~ x_7^*=1$ (table 14);
 $x_3^*=0 \Rightarrow x_1^*=1,~ x_2^*=0,~ x_4^*=1$ (table 13);
 $x_2^*=0 \Rightarrow x_5^*=1$ (table 12).
We found the optimal global solution to be $(1,~0,~0,~1,~1,~1,~1)$, the maximum objective
value is 18.
\end{example}
\section{Tree structural decompositions in discrete optimization}\label{tree_sec:5}
Tree structural decomposition methods use partitioning of
constraints and use as a meta-tree a structural graph .
Dynamic programming algorithm starts at the leaves of the meta-tree
and proceeds from the smaller
to the larger subproblems (corresponding to the subtrees) that is to say,
bottom-up in the rooted tree.
 \subsection{Tree decomposition and methods of its computing}
Aforementioned facts and an observation that many optimization
problems which are hard to solve on general graphs are easy on trees
make detection of tree structures in a graph a very promising
solution. It can be done with such powerful tool of the algorithmic graph theory
as a \textbf{tree decomposition} and the treewidth as a measure for the
''tree-likeness'' of the graph \cite{RobSey}. It is worth
noting that in \cite{Hlin08} is discussed a number of other useful parameters
like branch-width, rank-width (clique-width) or hypertree-width.
\begin{definition} Let $G=(X,E)$ be a graph. A
\textbf{tree decomposition} of $G$ is a pair $(T;\mathbf{Y})$ with $T = (I;
F)$ a tree and $\mathbf{Y} = \lbrace \mathbf{y_i} \mid I \in I \rbrace$ a family of
subsets of $X$, one for each node of $T$, such that
\begin{itemize}
\item (i) $\bigcup_{i \in I} \mathbf{y_i}=X,$ \item (ii) for every edge
$(x,y)\in X$ there is an $i\in I$ with $x \in \mathbf{y_i},~ y \in \mathbf{y_i}$,
\item (iii) (intersection property)  for all $i,j,l \in I$,
if $i<j<l$, then $\mathbf{y_i} \cap \mathbf{y_l} \subseteq \mathbf{y_j}$.
\end{itemize}
\end{definition}
Note that tree decomposition uses partition of
constraints, i.e., it can be considered as a dual structural decomposition method. The
best known complexity bounds are given by the ''treewidth'' $tw$
({\sc Robertson, Seymour} \cite{RobSey}) of an interaction graph
associated with a DOP. This parameter is related to some topological
properties of the interaction graph. Tree decomposition and the
treewidth ({\sc Robertson, Seymour} \cite{RobSey}) play a very
important role in algorithms, for many $NP$-complete problems on
graphs that are otherwise intractable become polynomial time
solvable when these graphs have a tree decomposition with restricted
maximal size of cliques (or have a bounded treewidth \cite{Arn91},
\cite{Bodl97}, \cite{BodKos07}). It leads to a time complexity in $O(n \cdot
2^{tw+1})$. Tree decomposition methods aim to merge
variables such that the meta-graph is a tree of meta-vertices.

The procedure to solve a DO problem with bounded treewidth
involves two steps: (1) computation of a good tree decomposition, and
(2) application of a dynamic programming algorithm that solves instances of
bounded treewidth in polynomial time.

Thus, a tree decomposition algorithm
can be  applied to solving DOPs using the following steps:
 \begin{enumerate}
   \item [(i)] generate the interaction graph for a DOP (P);
   \item [(ii)] using an ordering for Elimination Game add edges in the
   interaction graph to produce a (chordal) filled graph;
   \item [(iii)] build the elimination tree and information flows (see Fig 4(b));
   \item [(iv)] identify the maximum cliques, apply an absorption and build subproblems;
   \item [(v)]  produce a tree decomposition;
   \item [(vi)] solve the DO subproblems for each meta-node and combine
   the results using LEA.
 \end{enumerate}
As finding an optimal tree decomposition is $NP$-hard, approximate
optimal tree decompositions using triangulation of a given graph are
often exploited.
Let us list existing methods of computing tree decomposition using a survey
of them in \cite{Jegou_comp}. \textbf{ Optimal triangulations} algorithms
have an exponential time complexity.
Unfortunately, their implementations do not have much interest from
a practical viewpoint. For example, the algorithm described in
\cite{FomKT04} has time complexity  $O(n^4 \cdot (1.9601^n))$ \cite{Jegou_comp}.
 A paper \cite{GoDech04}
has shown that the algorithm proposed in \cite{ShoiGei97} cannot
solve small graphs (50 vertices and 100 edges). The
approach of \cite{GoDech04} using a branch and bound algorithm, seems promising for
computing optimal triangulations.
  \textbf{Approximation algorithms} approximate the optimum by a constant
factor. Although their complexity is often polynomial in the treewidth
\cite{Amir}, this approach seems unusable due to a big hidden constant.
 \textbf{ Minimal triangulation} computes a set $C^{'}$ such that,
for every subset $C^{''} \subset C^{'}$, the graph $G^{'} = (X,C
\cup C^{''})$ is not triangulated. The algorithms LEX-M \cite{RoTaLu76} and LB \cite{Berry99}
have a polynomial time complexity of $O(ne^{'})$ with $e^{'}$ the number of
edges in the triangulated graph.
  \textbf{Heuristic triangulation} methods  build a perfect elimination order
by adding some edges to the initial graph. They can be easily implemented and often do this
work in polynomial time without providing any minimality warranty.
In practice, these heuristics compute
triangulations reasonably close to the optimum \cite{Kjaerulff90}. \\
Experimental comparative study of  four triangulation algorithms,
LEX-M, LB, min-fill and MCS was done in  \cite{Jegou_comp}.
Min-fill orders the vertices from $1$ to $n$ by
choosing the vertex which leads to add a minimum number of edges
when completing the subgraph induced by its unnumbered neighbors.
Paper  \cite{Jegou_comp} claims that LB and min-fill obtain the best results.
\subsection{Computing tree decompositions for NSDP schemes}\label{sec_TD_constr}
Given a triangulated (or chordal) graph, the set of
its maximal cliques corresponds to the family of subsets associated
with a tree decomposition (so called \textbf{clique tree}
\cite{Blair93}). When we exploit tree decomposition, we only
consider approximations of optimal triangulations by clique trees.
Hence, the time complexity is then $O(n \cdot 2^{w^{+}+1})$ ($w+1
\leq w^{+}+1 \leq n$). The space complexity is $O(n \cdot s \cdot
2^{s})$ with $s$ the size of the largest minimal separator
\cite{Jegou_comp}.\\
Usually, tree decomposition is considered in the literature separately
from NSDP issues.
But there is a close connection between these two structural decomposition
approaches. Moreover, it is easy to see that a tree decomposition
can be obtained from the DAG of the computational NSDP procedure (this fact
 was noted  in \cite{KaDeLaDe05}).

Consider example \ref{primer} and build a tree decomposition associated
with the corresponding NSDP procedure. Associated
 underlying DAG of NSDP procedure for the variable ordering $\{x_5,x_2,x_1,x_4,x_3,x_6,x_7\}$
 is shown in Fig. \ref{InterGraph} (b).
 As was mentioned above, this underlying DAG of local variable
 elimination (the elimination tree) is constructed using
Elimination  Game. A node $i$ of the DAG is containing variables
$(\alpha_i, Nb_{G_{x_{i-1}}}^{(i-1)}(x_i))$ is
 linked with the first $x_j$ (accordingly to the ordering $\alpha$) which
 is in $Nb_{G_{x_{i-1}}^{(i-1)}}(x_i)$.
 Nodes and edges of desired tree decomposition correspond
 one-by-one to nodes  and edges of the underlying DAG.
 Each node of the tree decomposition is indeed
 a meta-node containing a subset of vertices of the
 interaction graph $G$. This subset induces a
 subgraph in $G$ that was condensed to generate the
 meta-node. Restore these subgraphs for each
 meta-node of the tree decomposition.
 \begin{proposition}  Graph structure obtained by this
 construction from the underlying DAG of the
 NSDP procedure is a tree decomposition.
\end{proposition}
Proof is in \cite{KaDeLaDe05}.\\
 In our example  \ref{primer}, we observe that the first (accordingly
 to ordering  $\alpha$) meta-node corresponds to the variable $x_5$
 and contains  variables (vertices) $x_2,x_5$ (i.e., $x_5 \cup Nb(x_5)$).
 Subgraph induced by these vertices can be constructed
 using the interaction graph $G$ (Fig. \ref{InterGraph} a). This subgraph
 is shown in Fig. 8 (a) --- the meta-node
 $\mathbf{y_1}$. Next meta-node $\mathbf{y_2}$ of the tree
 decomposition corresponds to the  variable $x_2$ and contains
 variables $x_1,~x_2,~x_3,~x_4$. The corresponding induced subgraph (clique)
 is shown inside the meta-node  $\mathbf{y_2}$ in Fig. 9 (a). Continuing in
 analogous way we obtain the tree  decomposition as shown in Fig. 8 (a). \\
 It is easy to see that some cliques in this tree decomposition
 are not maximal and can be absorbed by other cliques. In the case,
 when one clique contains  another clique, the second clique can
 be absorbed into the first one. Thus, the clique corresponding to
 the meta-node  $\mathbf{y_2}$ is absorbed by clique $\mathbf{y_3}$
 (we denote a result of absorption as a clique $\mathbf{y_{2,3}}$.
  The clique $\mathbf{y_5}$ is absorbed by clique $\mathbf{y_4}$ forming a clique $\mathbf{y_{4,5}}$.
  After absorptions done we obtain a clique tree (Fig. 8 (b)) containing only maximal cliques. These maximal
  cliques correspond to constraints of the DOP.
  In Fig. 8 (b)
maximal cliques and links between them are shown.
Local decomposition algorithm \cite{Shch83} that uses a dynamic programming paradigm
can be applied to this clique tree.
Other possible way of finding  the clique tree is using  maximal
spanning tree in the dual graph.\\
\subsection{Applying the local decomposition algorithm to solving DO problem}
To describe how tree decompositions are used to solve problems with
the local decomposition algorithm, let us assume we find a tree decomposition of a
graph $G$. Since this tree decomposition is represented as a rooted
tree $T$, the ancestor/descendant relation is well-defined. We can
associate to each meta-node $\mathbf{y}$ the subgraph of $G$ made up by the
vertices in $\mathbf{y}$ and all its descendants, and all the edges
between those vertices. Starting at the leaves of the tree $T$, one
computes information typically stored in a table, in a bottom-up
manner for each bag until we reach the root. This information is
sufficient to solve the subproblem for the corresponding subgraph.
To compute the table for a node of the tree decomposition, we only
need the information stored in the tables of the children (i.e.
direct descendants) of this node.  The DO
problem for the entire graph can then be solved with the information
stored in the table of the root of $T$.
Consider example \ref{primer} and exploit the tree decomposition (clique tree)
shown in Fig. 8 (b). Let us solve the subproblem corresponding
to the block $\mathbf{y_1}$. Since this
block is adjacent to the block $\mathbf{y_{2,3}}$, we have to solve a DOP with
variables $\mathbf{y_1}-\mathbf{y_{2,3}}$ for all possible assignments $\mathbf{y_1} \bigcap
\mathbf{y_{2,3}}$.
Thus, since $\mathbf{y_1}-\mathbf{y_{2,3}}=\left\{ x_5\right\}$ and $\mathbf{y_1}
\bigcap \mathbf{y_{2,3}}=\left\{ x_2 \right\}$, then induced subproblem has the
form:
\[ h_{\mathbf{y_1}}(x_2)=\max_{x_5}\left\{4x_5\right\} \]
subject to
\[
2x_2+3x_5 \le 4,~~x_j=0,1,~j \in \left\{2,5\right\}
\]
Solution of the problem can be written in a tabular form (see table 15).
\begin{center}\parbox[t]{40mm}{
 Table 15.\\ Calculation of $h_{\mathbf{y_1}}(x_2)$\\ \smallskip
\begin{tabular}{|c|c|c|} \hline
 $x_2$&   $h_{\mathbf{y_1}}$ & $x_5^*(x_2)$ \\
\hline 0 &   4 &   1 \\ 1 &   0 &   0 \\ \hline
\end{tabular}}
\qquad
\parbox[t]{50mm}{
Table 16. \\ Calculation of $h_{\mathbf{y_{2,3}}}(x_3)$\\
 \smallskip
 \begin{tabular}{|c|c|ccc|} \hline
 $x_3$ & $h_{\mathbf{y_{2,3}}}$ & $x_1^*(x_3)$ & $x_2^*(x_3)$ &  $x_4^*(x_3)$\\
\hline
0 &11&  1 &  0 & 1 \\
1 &6 &  1 &  0 & 0 \\ \hline
\end{tabular}}
\end{center} 
Since $\mathbf{y_{2,3}}-\mathbf{y_4}=\left\{ x_1,x_2,x_3,x_4\right\} - \left\{ x_3,x_6,x_7\right\}=\left\{ x_1,x_2,x_4\right\}$ and
$\mathbf{y_{2,3}}\bigcap \mathbf{y_4}=\left\{ x_3 \right\}$,
next subproblem corresponding to the leaf (or meta-node) $\mathbf{y_{2,3}}$ of the
clique tree is

\begin{center}
\includegraphics[scale=0.46]{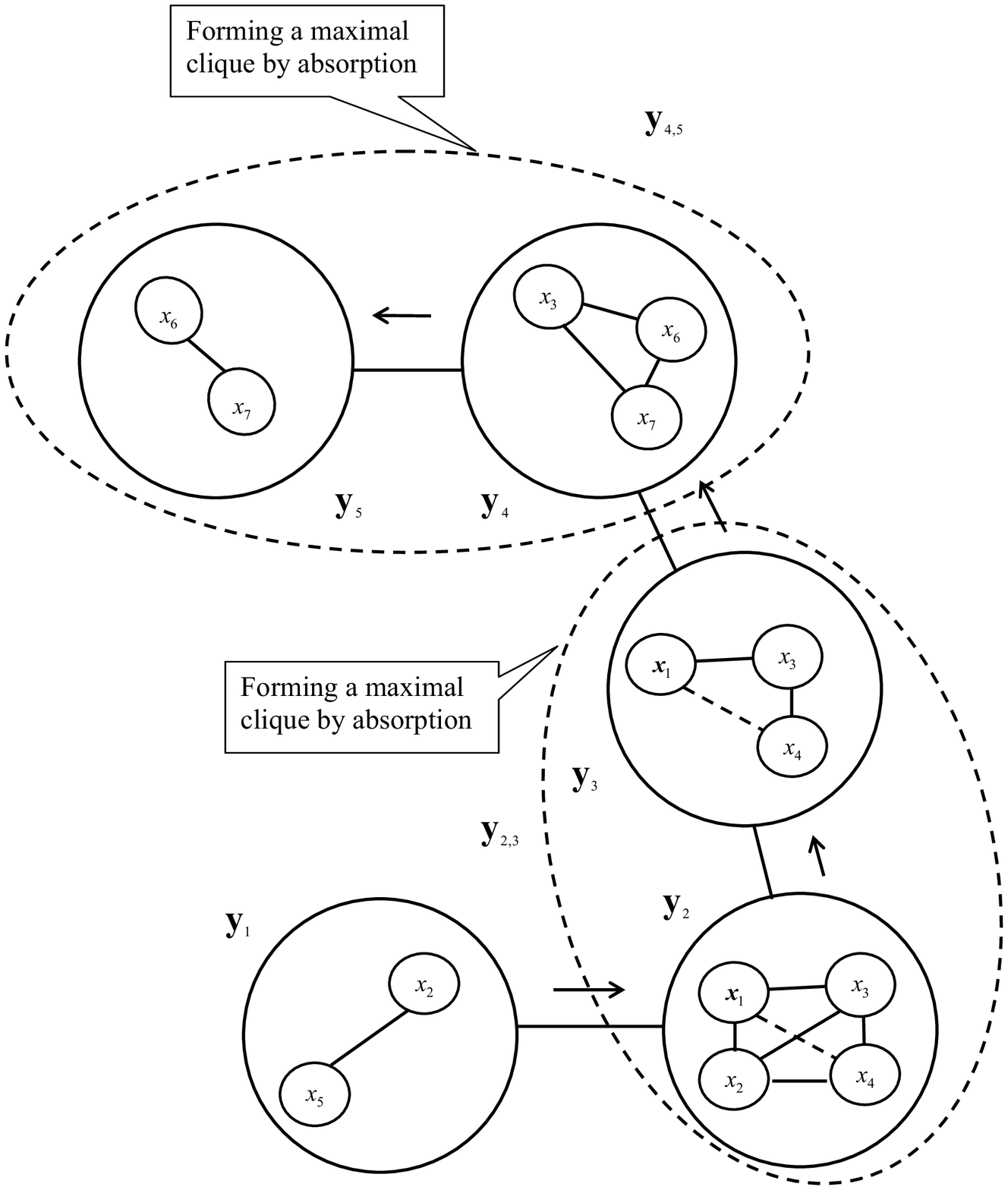}\\
(a)\\
\includegraphics[scale=0.46]{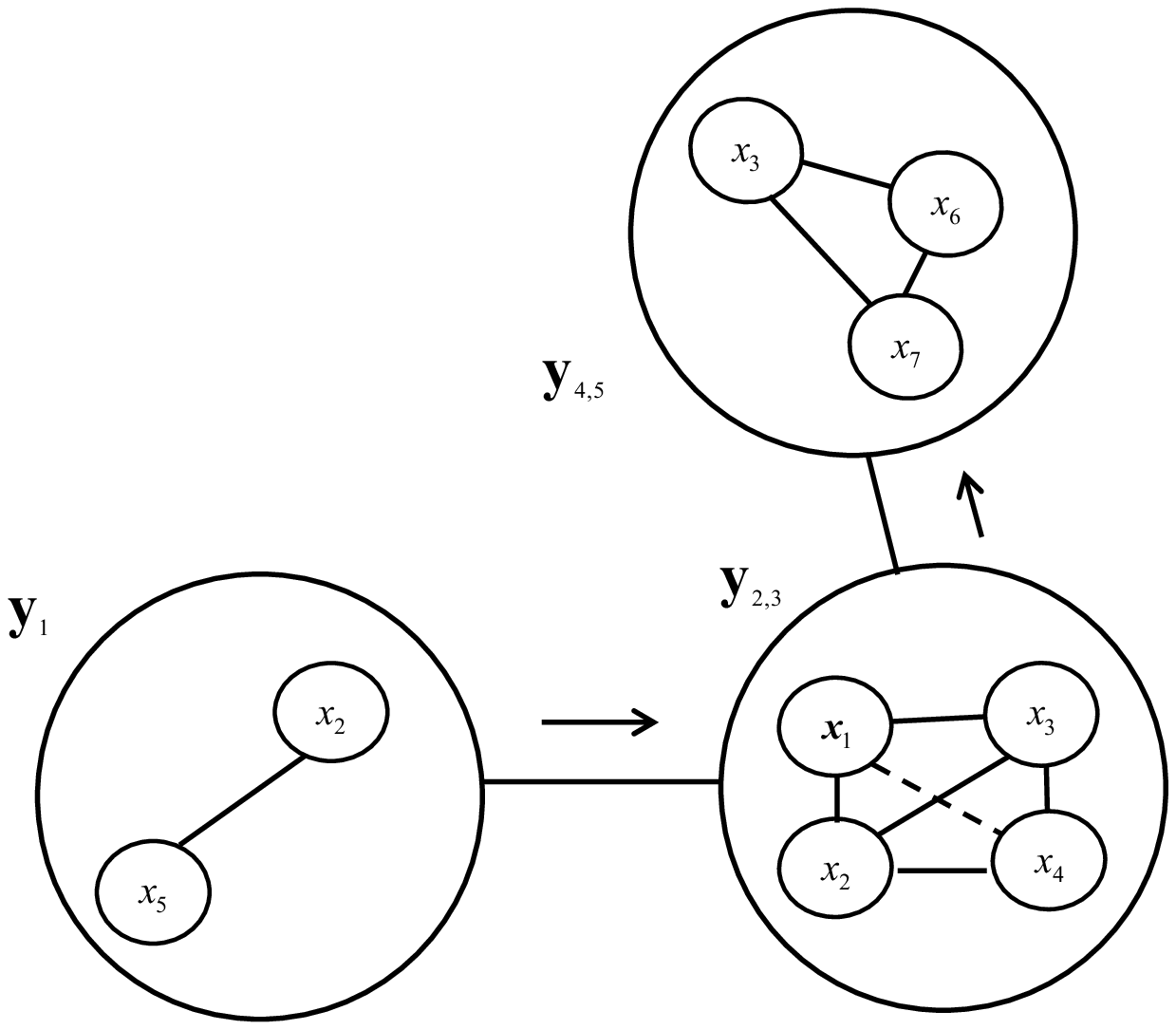}\\
(b)\\
Fig. 8. Tree decomposition for the NSDP procedure
(example \ref{primer}) before (a) and after absorption (b).
\end{center}

\[
h_{\mathbf{y_{2,3}}}(x_3)=\max_{x_1,x_2,x_4}\left\{h_{\mathbf{y_1}}+2x_1+3x_2+5x_4 \right\}
 \]
subject to
\[
3x_1+4x_2+x_3 \le 6,~~
2x_2+3x_3+3x_4 \le 5,
~~x_j=0,1,~j \in \left\{1,2,3,4 \right\}
\]
Solution of this subproblem is in table 16.
The last problem corresponding to the block $\mathbf{y_{4,5}}$ left to be solved is:
 \[
h_{\mathbf{y_{4,5}}}=\max_{x_3,x_6,x_7}\left\{h_{\mathbf{y_{2,3}}}(x_3)+x_3+6x_6+x_7 \right\} \] s.t.
\[ 2x_3+3x_6+2x_7 \le 5,~~x_j=0,1,~j \in \left\{3,6,7\right\}
\]
\begin{center} Table 17. Calculation of $h_{\mathbf{y_{4,5}}}$ \\
\label{t.table4}
\smallskip
\begin{tabular}{|c|ccc|} \hline
 $ h_{\mathbf{y_{4,5}}}$ & $x_3^*$ & $x_6^*$ & $x_7^*$ \\
\hline
18 &0&   1 &   1 \\ \hline
\end{tabular}
\end{center}
\bigskip
The maximal objective value is 18. To find the optimal values of the
variables, it is necessary to do a backward step of the dynamic
programming procedure: from table 17 we have $x_3^*=0,~ x_6^*=1,
~x_7^*=1$. From the table 16 using the information $x_3^*=0$
we find $ x_1^*=1,~x_2^*=0, ~x_4^*=1$. Considering table 15  we have for $x_2^*$=0:
 $x_5^*=1$. The solution is (1, 0, 0, 1, 1, 1, 1); the maximal objective
value is 18.

\section{Conclusion}
\label{sec:conc}
This paper reviews  the main graph-based local elimination algorithms for solving DO problems.
The main aim of this paper is to unify and clarify the notation
and algorithms of various structural DO decomposition approaches.
We hope that this will allow us to apply the aforementioned decomposition
techniques to develop competitive algorithms which will be able to solve
practical real-life discrete optimization problems.

%
%



\printindex
\end{document}